\documentclass[aps,prb,twocolumn,showpacs,preprintnumbers]{revtex4}
\usepackage{graphics}
\usepackage{graphicx}
\usepackage{dcolumn} 
\usepackage{bm}
\usepackage{color}
\usepackage{epsfig}
\usepackage{amsmath}
\newcommand{\koo}{KOsO$_4$}

\begin{document}
\title{Unquenched $e_g^1$ orbital moment in the Mott-insulating antiferromagnet KOsO$_4$
}
\author{Young-Joon Song$^1$, Kyo-Hoon Ahn$^1$}
\author{Kwan-Woo Lee$^{1,2}$}
\email{mckwan@korea.ac.kr}
\author{Warren E. Pickett$^{3}$}
\email{pickett@physics.ucdavis.edu}
\affiliation{
 $^1$Department of Applied Physics, Graduate School, Korea University, Sejong 339-700, Korea\\
 $^2$Department of Display and Semiconductor Physics, 
  Korea University, Sejong 339-700, Korea\\
 $^3$Department of Physics, University of California, Davis,                                          
  California 95616, USA
}
\date{\today}
\begin{abstract}
Applying the correlated electronic structure method based on density functional theory
plus the Hubbard $U$ interaction,
we have investigated the tetragonal scheelite structure Mott insulator KOsO$_4$, whose
$e_g^1$ configuration should be affected only slightly by spin-orbit couping (SOC). 
The method reproduces the observed antiferromagnetic Mott insulating state, populating
the Os $d_{z^2}$ majority orbital. 
The quarter-filled $e_g$ manifold is characterized by a symmetry breaking due to
the tetragonal structure, and the Os ion shows a crystal field splitting 
$\Delta_\emph{\emph{cf}}$ = 1.7 eV from the $t_{2g}$                                                                     
complex, which is relatively small considering the high formal oxidation state Os$^{7+}$.   
The small magnetocrystalline anisotropy before including correlation (i.e., in
the metallic state) is increased by more than an order of magnitude in the Mott-insulating
state, a result of a strong interplay between large SOC and a strong correlation. 
In contrast to conventional wisdom that the $e_g$ complex will not support orbital
magnetism, we find that 
for the easy axis [100] direction the substantial Os orbital moment $M_L\approx-0.2 \mu_B$
compensates half of the Os spin moment $M_S$ = 0.4$\mu_B$.
The origin of the orbital moment is analyzed and understood in terms of 
additional spin-orbital lowering of symmetry, and beyond that due to
structural distortion, for magnetization along [100]. Further
interpretation is assisted by analysis of the spin density and the
Wannier function with SOC included.
\end{abstract}
\maketitle

\section{Introduction}
In condensed matter, especially when containing heavy ions, 
spin-orbit coupling (SOC) leads to phenomena that are lacking without SOC. Examples
of recent interest include the original
topological insulators,\cite{Kane} behavior arising from the Rashba effect, 
unconventional metal-insulating transitions, 
compensating spin and orbital moments,\cite{bnoo,meetei}
and the magnetocrystalline anisotropy (MCA) that is so important in spintronics applications.
Whereas SOC within a $t_{2g}$ manifold in a ${M}$O$_6$ octahedron 
(${M}$ = transition metal) has a long history\cite{stevens,enough,lacroix} and 
has been intensively discussed recently in several specific
contexts,\cite{bnoo,sr2iro4,balents,dodds,naoso3,mat} 
corresponding effects in an $e_g$ manifold have rarely been considered due to
the conventional wisdom that the $e_g$ subshell ensures a
perfectly quenched orbital moment.
From this viewpoint, heavy transition metal oxides containing ${M}$O$_4$ tetrahedra 
are of great interest, since crystal field splitting leads to partially filled orbitals
in the $e_g$ manifold.

About a century ago, monoclinic crystals of two toxic, volatile materials, OsO$_4$ and RuO$_4$,
were synthesized.
These are presumably textbook band insulators, albeit with remarkably high (8+)
formal charges. 
Although existing data  on these crystals are limited, the effects of SOC
have been investigated from a chemical viewpoint since the 1990's\cite{perez,persh1,persh2}
and have been generally found to be minor.
In 1985, heptavalent ${A}$OsO$_4$ (${A}$ = alkali metal) compounds were synthesized by
Levason {\it et al.}, who determined they formed in the tetragonal scheelite 
crystal structure.\cite{levason}
\koo~ has been often synthesized from a mixture of KO$_2$ and Os metal 
as a precursor for preparation of
the superconductor KOs$_2$O$_6$,\cite{hiroi}
but further investigations of its physical properties are still lacking.
\koo~ seems to be insulating, though detailed resistivity
data are not yet available.\cite{koda,yamaura1}

Recently, Yamaura and collaborators determined the crystal structure parameters and measured the
susceptibility and specific heat.\cite{yamaura1}
The Curie-Weiss moment is 
$\mu_\emph{\emph{eff}}$ = 1.44$\mu_B$, 20\% reduced                                                         
from the spin-only moment, and the N\'eel temperature is $T_N$ = 37 K.                             
These authors suggested that magnetic frustration in this distorted diamond lattice may be
necessary to account for observations. However, the conventional ratio of
Curie-Weiss to ordering temperatures  $|\theta_{CW}|/T_N \approx$ 1.8 is small (i.e.,
there is little frustration in the bipartite Os sublattice) so other factors must be considered.

In this paper we study the electronic structure of KOsO$_4$, with special attention
given to the interplay between strong correlation and SOC. The small ligand field splitting
of the $e_g$ orbitals due to distortion of the OsO$_4$ tetrahedron plays an important
role in determining the occupied orbital in the Mott-insulating state, and may become
active in effects arising from SOC as well. A modest $t_{2g}$-$e_g$ crystal field
splitting ($\Delta_\emph{\emph{cf}}$ = 1.7 eV) and large SOC strength ($\sim$0.3 eV) bring in another         
effect of crystallinity that impacts the effects of SOC. This splitting is especially
small considering that in another Os$^{7+}$ compound, the double
perovskite Ba$_2$NaOsO$_6$, $\Delta_\emph{\emph{cf}}$                                                         
= 6 eV is extremely large.\cite{bnoo} Results are analyzed in terms
of magnetization densities, Wannier functions, and spin-orbital occupation numbers.
Symmetry reduction of the electronic state due to SOC when the spin lies in the 
[100] direction is found
to have a great consequence:
A population imbalance of the $m_l=\pm2$ orbitals 
leads to an unexpectedly large orbital moment, as discussed in Sec. V.

\begin{figure}[tbp]
{\resizebox{7.5cm}{6.5cm}{\includegraphics{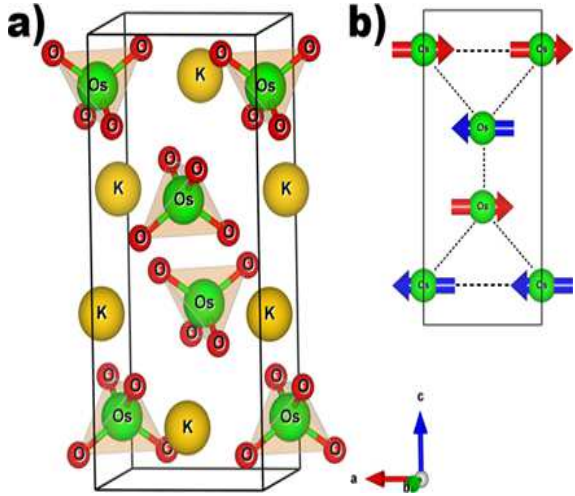}}}
\caption{(Color online) (a) Scheelite-type crystal structure of \koo.
(b) $G$-type antiferromagnetic (AFM) spin ordering, which is the ground state
in LSDA+$U$+SOC calculations. The arrows indicate the calculated directions of spins (easy axis).
}
\label{str}
\end{figure}

\section{Structure and calculation methods}
\koo~ crystallizes in the scheelitelike structure (space group: $I4_1/a$, No. 88), shown in Fig. \ref{str}.  
In this tetragonal structure with two formula units (f.u.) per primitive cell, 
the lattice parameters are $a$ = 5.652 \AA~ and $c$ = 12.664 \AA,\cite{koo}
leading to a ratio of $c/\sqrt{2}a = 1.58$. The Os sublattice forms a
substantially  elongated
diamond sublattice; this  $c/\sqrt{2}a$ ratio is unity for the cubic 
diamond lattice.
The K and Os atoms sit at the $4b$ sites (0,$\frac{1}{4}$,$\frac{5}{8}$) and 
$4a$ sites (0,$\frac{1}{4}$,$\frac{1}{8}$), respectively.
The O atoms lie on the $16f$ sites (0.1320,0.0160,0.2028).
In the OsO$_4$ tetrahedron, all Os-O bond lengths are 1.81 \AA, 
and the O-Os-O bond angles are either 114$^\circ$ or 107$^\circ$, 
compared to 109.5$^\circ$ for a regular tetrahedron.
A similar distortion is observed in the band insulator OsO$_4$,\cite{oso4} 
while both RuO$_4$ and KRuO$_4$ have nearly ideal tetrahedra.\cite{ruo4,kruo4}
This difference suggests that the distortion is due to a chemical difference between Os
and Ru ions.

Our calculations were carried out with the local (spin) density approximation [L(S)DA]
and its extensions, as implemented in the accurate all-electron full-potential code 
{\sc wien\small{2}\normalsize{k}}.\cite{wien2k}
Since we are interested in a possible competition between large SOC and strong
correlation effects in magnetic systems, we compare all of the LDA, LSDA,
LSDA+SOC, LSDA+$U$, and LSDA+$U$+SOC approaches.
An effective on-site Coulomb repulsion $U$ was used for the LDA+$U$ calculations; since Os$^{7+}$
is a $d^1$ ion which is not occupied by more than one electron, the Hund's rule coupling
$J$ between two electrons of the same spin was set to zero.
To analyze the partially filled Os complex, the Wannier function approach 
implemented in {\sc fplo} and {\sc wien\small{2}\normalsize{k}} has been used.\cite{fplo1,kunes}                       
Calculations of the Wannier function including SOC are available only in the latter.
In {\sc wien\small{2}\normalsize{k}}, the following muffin-tin radii are adopted: 2.02 for Os, 1.4 for O, and 2.2 for K 
(in units of a.u.). The extent of the basis was determined by $R_\emph{\emph{mt}}K_\emph{\emph{max}}$ = 7.                    
The Brillouin zone was sampled with a sufficiently dense $k$ mesh (for
an insulator) of $13\times 13\times 6$.

\begin{figure}[tbp]
\vskip 8mm
{\resizebox{7.5cm}{6cm}{\includegraphics{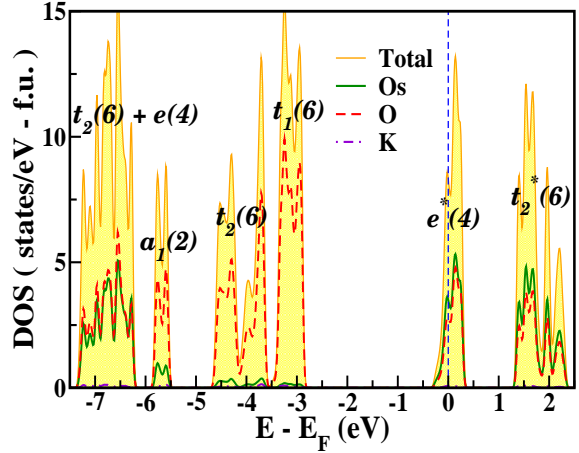}}}
\caption{(Color online) LDA total and atom-projected densities of states (DOS)
of nonmagnetic \koo~ 
in the regimes of Os $5d$ and O $2p$ orbitals.
The symbols, which are displayed in each manifold, represent the molecular orbitals of 
the OsO$_4$ tetrahedron, following the notations of Ref. [\onlinecite{will}].
The values in parentheses indicate the number of bands in each manifold.
The symbol $\ast$ denotes the antibonding state.
The DOS $N(E_F)$ at the Fermi energy $E_F$, which is set to zero, is 
4.18 states/eV f.u. spin.
}
\label{nm_dos}
\end{figure}

\begin{figure}[tbp]
\vskip 8mm
{\resizebox{7cm}{6cm}{\includegraphics{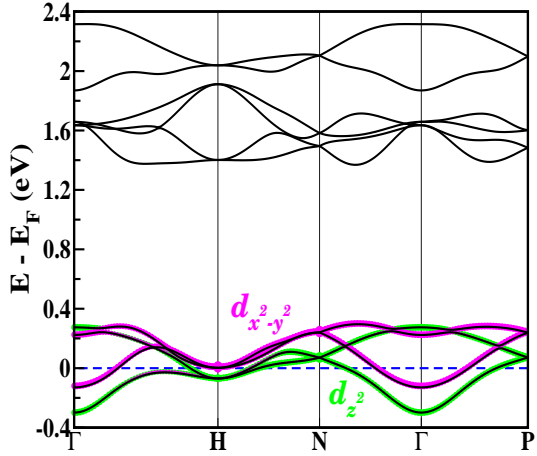}}}
\caption{(Color online) LDA Os $5d$ band structure of nonmagnetic \koo,
showing an $e_{g}$--$t_{2g}$ crystal field splitting of $\sim$1.8 eV.
The partially filled $e_g$ bands, which are colored with the corresponding 
Wannier orbitals, lie on the range of --0.3 to 0.3 eV.
In units of ($\pi/a$,$\pi/a$,$\pi/c$), the symmetry points shown
are $H=(100)$, $N=(\frac{1}{2}\frac{1}{2}0)$, 
and $P=(\frac{1}{2}\frac{1}{2}\frac{1}{2})$.
}
\label{nm_band}
\end{figure}

\section{The underlying electronic structure}
Figure \ref{nm_dos} displays the LDA total and atom-projected densities of states 
(DOSs), which demonstrates a strong $p$-$d$ hybridization not only in the
most relevant Os $e_g$ bands (denoted as the molecular $e^*$ orbitals)
but also in more tightly bound oxygen orbitals around --7 eV. This
hybridization of the transition metal $d$ character into O $2p$ bands is common but is not particularly
relevant and is little discussed.
The narrow bands reflect moderately banding
molecular orbitals.
Some nearly pure oxygen bands lie in the --6 to --3 eV range.
The $t_2^*$ bands centered around 2 eV are a mixture of Os $t_{2g}$, 
and all O $2p$ orbitals,
while the $e^*$ set is a mixture of $e_g$ and mostly $p_\pi$.

Before considering the complications of spin polarization, correlation effects, and SOC,
we consider the basic underlying features of the electronic structure.
Supposing formal charges of K$^{+}$, Os$^{7+}$, and O$^{2-}$ ions, 
the crystal field $e_{g}$--$t_{2g}$ splitting is expected to be 0.8 eV, about
half of the calculated splitting $\Delta_\emph{\emph{cf}}$ = 1.7 eV,                                           
which is the full ligand field splitting. 

The LDA nonmagnetic band structure in the Os $5d$ band region (ten bands due to 2 f.u. 
per primitive cell)
is displayed in Fig. \ref{nm_band}. 
The distortion of the OsO$_4$ tetrahedron leads to the crystal field splitting
of $d_{xy}$ above the degenerate pair \{$d_{xz}$,$d_{yz}$\}, 
as is evident in Fig. \ref{nm_band}.
Notably, the isolated partially filled $e_g$ manifold can be fit well 
using an effective two-band model with three nearest neighbor hopping parameters.
The hopping parameters corresponding to the corresponding Wannier functions are 
\begin{align}
t_{1}&=&\langle d_{x^2-y^2}&\mid\hat{H}\mid d_{x^2-y^2}\rangle&=&\rm{43 ~meV},\nonumber \\
t_{2}&=&\langle d_{z^2}&\mid\hat{H}\mid d_{z^2}\rangle        &=&\rm{56 ~meV}, \nonumber \\
t'   &=&\langle d_{z^2}&\mid\hat{H}\mid d_{x^2-y^2}\rangle    &=&~~\rm{7~ meV}.
\end{align}
The site energies are 59 meV for $d_{z^2}$ and 143 meV for $d_{x^2-y^2}$ relative
to $E_F$. It is this ligand field splitting of 84 meV
that determines that the $d_{z^2}$ becomes
occupied in the Mott-insulating phase (below).
As expected from the small value of $t'$, each of the $d_{x^2-y^2}$
and $d_{z^2}$ bands can be fit nearly as well along symmetry lines
using two independent single band models.
A noticeable mixing between the two bands only occurs along the $\Gamma$-$H$ line.
The superexchange coupling parameter is determined from $J=t^2/U \sim$ 2 meV,
using $U$ = 2 eV (see below). The magnitude of this exchange coupling is similar
to the ordering temperature $k_B$$T_N$ $\approx$ 3 meV. 

\begin{figure}[tbp]
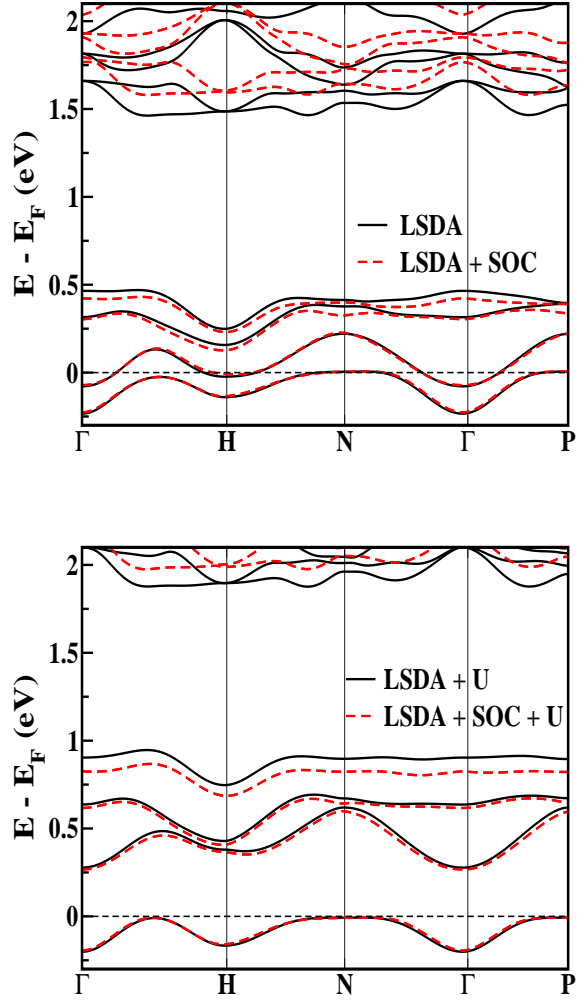

\vskip 8mm
{\resizebox{7.5cm}{6cm}{\includegraphics{Fig4a.eps}}}
\vskip 12mm
{\resizebox{7.5cm}{6cm}{\includegraphics{Fig4b.eps}}}
\caption{(Color online) AFM band structures of (top) LSDA without and with SOC, and
(bottom) LSDA+$U$ without and with SOC, for $U=2$ eV. 
In the insulating state,  
the occupied state is mainly $d_{z^2}$. 
}\label{afm_band}
\end{figure}  

\begin{figure}[tbp]
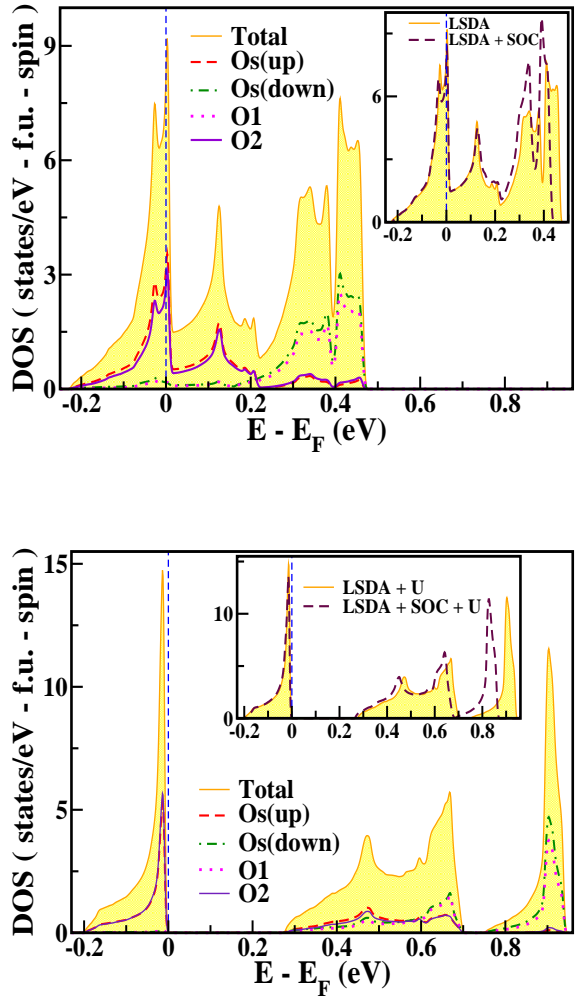

\vskip 8mm
{\resizebox{7.5cm}{6cm}{\includegraphics{Fig5a.eps}}}
\vskip 12mm
{\resizebox{7.5cm}{6cm}{\includegraphics{Fig5b.eps}}}
\caption{(Color online) AFM densities of states of (top) LSDA and (bottom) LSDA+$U$
at $U$ = 2 eV, with atomic contributions differentiated.  
{\it Inset}: Comparison of DOS with the case including SOC, near $E_F$.
In the metallic $U$ = 0 state, the exchange splitting in the $e_g$ manifold is 0.4 eV.
In the insulating state, in terms of the spin-up Os,
the plots contain the filled spin up $d_{z^2}$, the unfilled $d_{x^2-y^2}$, 
and the unfilled spin down $d_{z^2}$ bands, from the lower energy.
In LSDA, $N(E_F)\approx 4$ states/eV spin f.u., but lies on a very sharp edge.
Inclusion of SOC reduces $N(E_F)$ by 7\%.}
\label{afm_dos}
\end{figure}

\begin{table*}[bt]
\caption{Effect of correlation $U$ on the relative energies $\Delta E$ 
(in units of meV/f.u.) and Os orbital moments 
 $M_{L}$ (in units of $\mu_B$) for each of four spin quantization directions
and for FM and AFM alignments.
$M_L$ of Os is antialigned to the spin moment of Os, 
which is $\sim$0.4$\mu_B$ for the insulating states. 
The spin moments contributed by O ions are 0.24 -- 0.32$\mu_B$/4O in LSDA+SOC, 
increasing to $\sim$0.4$\mu_B$/4O for LSDA+SOC+U.
$U$ = 2 eV was used for LSDA+SOC+$U$ calculations.
}
\begin{center}
\begin{tabular}{ccccccccccc}\hline\hline
 ~ & ~    & \multicolumn{4}{c}{AFM} &~ & \multicolumn{4}{c}{FM}\\\cline{3-6}\cline{8-11}           
 ~&~  & ~[100]~& ~[001]~&~[110]~&~[111]~& ~& ~[100]~&~[001]~&~[110]~&~[111] \\\hline
LSDA+SOC   & $\Delta E$ & 0 & 4.6 & 2.3 & 3.7 & ~& 1.9 & 3.9 & 1.8 & 3.4\\
LSDA+$U$+SOC & $\Delta E$ & 0 & 14.4& 3.0 & 10.6 & ~& 19.3 & 28.7 & 19.8 & 26.2 \\
LSDA+SOC   & $M_{L}$    & --0.134 & --0.014 & --0.136 & --0.052 & ~& --0.135 & --0.048 &--0.135&--0.073\\
LSDA+$U$+SOC & $M_{L}$    &~--0.184~&~0.007~&~--0.183~&~--0.053~& ~&~--0.176~&~0.006~&~--0.172~&~--0.055 \\\hline
\end{tabular}
\end{center}
\label{table1}
\end{table*}

\section{Effects of correlation and SOC}
A primary emphasis in our study of this system is to assess the interplay
in an $e_g$ system between
strong correlations, which prefer full occupation of certain orbitals and usually increase
spin polarization, and SOC, which mixes spin orbitals and complicates all aspects of
the electronic structure while inducing the orbital moment and magnetocrystalline 
anisotropy (MCA).
It was mentioned above that including correlation effects in the LSDA+$U$ method
leads to preferred occupation
of the $d_{z^2}$ orbital, which has 84 meV lower on-site (crystal field)
energy than $d_{x^2-y^2}$ due to
the distortion of the OsO$_4$ tetrahedron.
The band structures including the lower part of the $t_{2g}$ complex, and the DOS
of the $e_g$ bands alone for the energetically preferred AFM state, are displayed
in Figs. \ref{afm_band} and \ref{afm_dos}, respectively. These figures have been
constructed to allow identification of the individual effects
of $U$ and SOC. 

Before proceeding with a description of the full electronic structure and then the
spin density itself, we review the energy differences arising from the various
interactions.

\subsection{Magnetic Energy Differences}
As expected from the peak at $E_F$ in the DOS (see Fig. \ref{nm_dos}) and 
the well known Stoner instability, 
ferromagnetism (FM) is energetically favored over the nonmagnetic state, by 26 meV/f.u. 
Our fixed spin moment calculations of the interacting susceptibility\cite{krasko1987} 
lead to $IN(E_F)$ = 1.60, well
above the Stoner instability criterion of unity, 
and $N(E_F)$ = 4.09 states/eV f.u. spin gives the
Stoner parameter $I$ = 0.39 eV, similar to the value obtained\cite{bnoo} 
for Ba$_2$NaOsO$_6$.
Within metallic LSDA where exchange coupling might be considered to be some mixture of
double exchange, Ruderman-Kittel-Kasuya-Yosida (RKKY), and superexchange, 
the FM ground state lies 5.5 meV/f.u. below the observed AFM state.

To assess the effects of SOC before including correlation corrections, 
we display MCA energies with LSDA+SOC with several spin 
orientations in Table \ref{table1}.
[100] is the AFM easy axis, however, all spin orientations differ little in energy compared
to the larger differences when all interactions are included (see below),
so the magnetic anisotropy is predicted to be small at this level of 
theory. This tentative conclusion, before including correlation, is consistent with conventional
wisdom that SOC has little effect in $e_g$ systems.

After including correlation with $U$ = 2 eV, the AFM Mott-insulating state is obtained
(discussed below) as the ground state, by 19 meV over FM alignment.
This favoring of antiferromagnetism over ferromagnetism 
is common when applying the LDA+$U$ functional in transition metal oxides.
For bipartite AFM (alternating) alignment compared to FM alignment, 
the AFM magnetic coupling is $J\approx$ 4.8 meV $\approx$ 56 K, 
consistent in magnitude with the experimental ordering temperature $T_N$ = 37 K.

Other aspects of the interplay between strong SOC and strong correlation are apparent
in Table \ref{table1}. Most notable in the energetics is that strong correlation
effects (i.e., including $U$) greatly enhance the MCA: Energy differences
between different directions of the spin are more than an order of magnitude larger. 
This is more surprising when one recalls that SOC effects (which provide the
MCA) are often supposed to be negligible in $e_g$ subshells. Including both large $U$ and
SOC, the [100] direction is now very clearly determined as the easy axis.

\begin{figure}[tbp]
\vskip 8mm
{\resizebox{8cm}{4cm}{\includegraphics{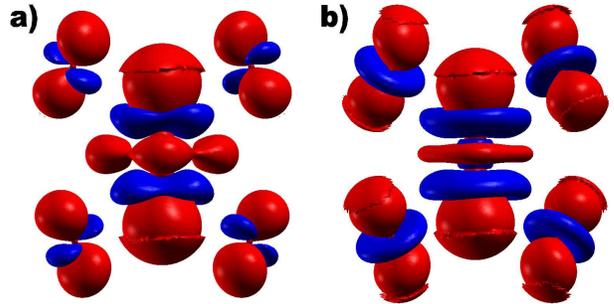}}}
\caption{(Color online) Isosurface plots at 0.042 $e$/\AA$^3$ of spin densities
for (a) metallic ($U$ = 0) and (b) insulating AFM ($U$ = 2 eV) \koo,
when including SOC. Os at the center is surrounded by four O ions.
Red indicates majority spin, and blue denotes minority.
The insulating density in (b) reflects the circular shape around
both Os and O that provides the orbital moment.
}
\label{CD}
\end{figure}

\subsection{LSDA+SOC+$U$ leads to a Mott-insulating state}
The lowest and highest bands (Fig. \ref{afm_band})
in the $e_g$ manifold extending over the regime of --0.25 to + 0.5 eV
are the Os spin-up and spin-down $d_{z^2}$ bands, respectively.
In the quarter-filled $e_g$ manifold, the $d_{z^2}$-$d_{x^2-y^2}$ degeneracy lifting
is 0.2 eV, i.e., the $e_g$ degeneracy is already split (presumably self-consistently
by occupation of the $d_{z^2}$ orbital and the resulting Jahn-Teller distortion).
Applying the on-site Coulomb repulsion $U$ starting from small values
leads to a metal-insulator transition (MIT) (gap opening) at a critical value $U_c \approx$ 1.2 eV,
which is near the bottom of the range of expected values for Os.
As shown in the bottom panel of Fig. \ref{afm_band} for $U$ = 2 eV and spin along
the [100] direction, the top of the occupied band
has a flat region around the $N$ point, giving rise to a one-dimension-like peak
and sharp discontinuity in the DOS at the top of the band, evident in Fig. \ref{afm_dos}.
Other band maxima at $P$ and midway between $\Gamma$ and $N$ are (somewhat accidentally)
degenerate with the flat band at $N$.
The occupied bandwidth is 0.2 eV.
As shown by the red dashed lines in the band structure of Fig. \ref{afm_band},
inclusion of SOC a has negligible effect on the occupied state (position and
dispersion) but lowers the uppermost $e_g$ band (primarily minority spin) by 0.15 eV.
This shift corresponds to a small decrease in the exchange splitting of the
unoccupied $e_g$ orbital.

\subsection{Effects of SOC on spin and orbital moments}
In the following text and in Table \ref{table1} we quote atomic moments
from contributions within the inscribed spheres, which are
somewhat smaller than the full value. We remind that the occupied ``$d_{z^2}$'' orbital
that is occupied before including SOC is strongly hybridized with $2p$ orbitals
of the surrounding O ions, so the spin magnetization of 1$\mu_B$ is distributed over
oxygen as well as Os. The moment values should be considered in conjunction with the
spin density isosurfaces pictured in Fig. \ref{CD}.

For all spin orientations we have determined that the Os spin moment is
$M_S \approx$ 0.4$\mu_B$. This value is almost independent of $U$ in the range
0--5 eV that we have studied.
The O net spin $M_S$ = 0.07$\mu_B$/O aligns parallel to
that of the nearest neighbor Os. The sum of the full atom moments must be unity,
so atomic values are around 0.5$\mu_B$ and 0.12$\mu_B$ for Os and O, respectively,
versus the atomic sphere values just quoted.
Including SOC reduces the Os spin moment by 10\%, transferring the
difference to neighboring O $M_S$
due to rehybridization. Nonzero $M_L$ must arise from mixing in of $t_{2g}$
character, as we discuss in Sec. V.
For [100] and [110] spin directions, increasing $U$ increases $|M_L|$ from $\sim$1/3$M_S$ 
at $U$ = 0 to 1/2$M_S$ at $U$ = 2 eV.
For [001] spin orientation $M_L$ is essentially vanishing for any value of $U$
(Table \ref{table1}).

\subsection{Behavior of the spin density}
The spin density isosurface plots displayed in Fig. \ref{CD} for metallic
($U$ = 0) and AFM insulating ($U$ = 2 eV) phases are instructive.
Even in the metallic, uncorrelated case both positive and negative lobes of spin density
appear on both Os and O ions, indicating more complexity than strong (but
typically simple) $p$-$d$ hybridization. Since in this limit the lower Hubbard band is fully
polarized (only spin-up states), the negative polarization arises from polarization
within the filled nominally O $2p$ bands at lower energy.

The net spin of O lies in $2p$ orbitals whose orientation reflects a $\pi$
antibonding character with Os $d_{xz}, d_{yz}$ orbitals.
A small negative spin density is
induced in a linear combination of the $p_x$, $p_y$ orbitals, in the local coordinate system. 
As expected, the Os spin lies mainly in the $d_{z^2}$ orbital, with some admixture
of $d_{x^2-y^2}$ accounting for the square versus circular symmetry of
the spin density in the equatorial plane of Os.
A small but clear admixture of
$d_{yz}$ and $d_{xz}$ character appears as a negative spin density (blue), and
this contribution is
necessary to provide the Os orbital moment.

In the correlated ($U$ = 2 eV) insulating state, Os still
 has mainly a $d_{z^2}$ character for spin up.
However, the circular symmetry indicating $d_{yz}$-$id_{xz}$ character for spin down
shows up much more clearly.
Unexpectedly, this same development of $p_x$-$ip_y$ (in an appropriate local
frame) shows up on the O ions in the
spin-down region, while the spin-up, local $p_x$ character is nearly undisturbed.

\begin{figure}[tbp]
\vskip 8mm
{\resizebox{7cm}{5cm}{\includegraphics{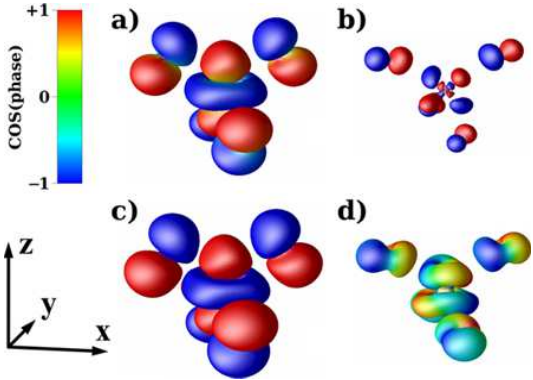}}}
\caption{(Color online) 
Isosurface plots of the Wannier function density $|W_{\sigma}(r)|^2$
of the occupied band in AFM insulating \koo
(LSDA+$U$+SOC with $U$ = 2 eV), for the majority $\sigma=|\uparrow\rangle$  
and minority $\sigma=|\downarrow\rangle$ spins separately.  
The surfaces are colored with the cosine of the phase of each component
(real positive, red; real negative, blue; imaginary. green), as described in
color legend bar.
(a), (b) Spin in the [100] direction;  
(c), (d) spin in the [001] direction.
(a) and (c) are for the majority spin, shown at isovalue = 2 a.u.
(b) and (d) are for the minority spin, shown at the smaller isovalue of 0.3 a.u.  
While the minority spin component is small, its directional dependence is
evident, with a much larger imaginary part for [001].
}
\label{WF}
\end{figure}

To generate the complex-valued, mixed-spin Wannier function $W(r)$ of the occupied band,
we projected from $|\frac{5}{2},\frac{1}{2}\rangle$ and $|\frac{3}{2},\frac{1}{2}\rangle$
as a trial function in the {\sc wien\small{2}\normalsize{wannier}} package.                                  
Figure \ref{WF} presents isosurface plots of $|W(r)|^2$ for each of the two components
of spin.
The spin-down part is much smaller than the spin-up part,
as indicated separately by the spin moment which remains close to 1$\mu_B$/f.u.
Consistent with Eqs. (\ref{eq2}) and (\ref{eq3}) below, the spin-down parts are
$d_{xz}$-like in the [100] direction and $d_{xz}$-$id_{yz}$-like
in the [001] direction.
The spin-up parts are $d_{z^2}$-like in both directions,
but have a squarish negative lobe rather than a circular shape.
The complex character of the spin-up part in the [100] direction is visible only
around the neck of the positive lobe, since the imaginary part is small.
However, this complex character leads to a symmetry breaking between
$m_l=\pm2$ orbitals,
as will be discussed below.
The spin-up part in the [001] direction is purely real.

\section{Analysis of spin-orbit coupling in the $e_g^1$ case}

Now we address the effects of SOC, especially the appearance of a surprisingly large
orbital moment in an $e_g$ subshell which should not produce an orbital moment, 
through analysis of the occupation matrices 
and the associated Wannier function.
SOC effects in the $e_g$ channel tend to be relegated to the
background because $e_g$ contains only
orbital $m_l = 0$ and $m_l = \pm2$ $d$ orbitals, which are not coupled by the electron
spin $s$ = $\frac{1}{2}$. Note, however, that this is strictly true only in the spherical
(isolated ion) limit and for orbital moments along the axis of quantization,
i.e., the direction of the spin. Indeed, we find negligible orbital moments for spin
along [001]. Crystalline effects break this orbital-moment
killing symmetry.

First, in KOsO$_4$, the crystal field splitting $\Delta_\emph{\emph{cf}} = 1.7$ eV                    
is slightly larger than the
SOC strength, so virtual inclusion of $t_{2g}$ orbitals may be involved. 
Second, the OsO$_4$ tetrahedron is distorted, breaking the twofold $e_g$ symmetry, which is
related to the Mott-insulating behavior: occupation of a single $e_g$ orbital and the
accompanying Jahn-Teller distortion. Finally,
the higher symmetry crystalline $\hat z$ axis is not the easy axis, so additional
complexities arise. 
The focus begins with the $d_{z^2}$ orbital that is occupied before SOC is included,
with a slight admixture of other $5d$ orbitals due to structural symmetry
breaking and hopping.

{\it Spin along [001].} Applying $\vec{L}\cdot\vec{S}$ to the spin-up 
$d_{z^2}$ orbital leads to
\begin{eqnarray}
\vec{L}\cdot\vec{S}|d_{z^2}\rangle|\uparrow\rangle_z\propto
                   -(|d_{xz}\rangle +i|d_{yz}\rangle)|\downarrow\rangle_z,
\label{eq2}
\end{eqnarray}
which are nominally unoccupied orbitals. Indeed,
we calculate negligible $M_L$ for this orientation, reflecting negligible intermixing
of $d_{xz} \pm i d_{yz}$ orbitals across the crystal field gap $\Delta_\emph{\emph{cf}}$.             
The main occupation amplitudes (eigenvectors of the occupation matrix)
are (in $|m_l,m_S\rangle$ notation), 0.96 for $|0,\uparrow\rangle$                               
and a down-spin amplitude of --0.21 for $|+1,\downarrow\rangle$ (thus decreasing the        
spin moment by 4\%).

{\it Spin along [100]}. For in-plane  [100] spin orientation SOC leads to 
the common picture
\begin{eqnarray}
\vec{L}\cdot\vec{S}|d_{z^2}\rangle|\uparrow\rangle_x
             \propto-i|d_{yz}\rangle|\uparrow\rangle_x
             -|d_{xz}\rangle|\downarrow\rangle_x.
\label{eq3}
\end{eqnarray}
Another way to approach the emergence of an orbital moment is to note that
when the $d_{z^2}$ orbital is expressed in local coordinate system
$X,Y,Z$, with $Z$ directed along [100],  it is a linear combination of
$d_{Z^2}$ and $d_{X^2-Y^2}$ orbitals (i.e., the $e_g$ orbitals). 
Breaking of symmetries may induce an asymmetry in the $m_l=\pm 2$ orbitals
making up $d_{X^2-Y^2}$. Indeed, this happens. 
Table \ref{table2} shows the amplitudes of the occupation matrix eigenstate 
in the local coordinate system.
The imbalance in the $m_l=+2$ and $m_l=-2$ occupations in the spin-up channel
results in a surprisingly large (for an $e_g$ shell) orbital moment.

\begin{table}[bt]
\caption{Amplitude coefficients of the occupied orbital, expressed
with respect to complex orbitals and both spin components,
in the the local coordinate system 
with spin along the [100] direction.
}
\begin{center}
\begin{tabular}{ccccccc}\hline\hline
 ~ & ~ & \multicolumn{5}{c}{$m_l$} \\\cline{3-7}
  ~&~  &~~ 0~~ &~~--1~~&~~+1~~& ~~ --2~~& ~~ 2~~ \\\hline
 $|\uparrow\rangle$   &~~& --0.47 &~~ --0.09$i$&~~ $0.11i$ &~~ 0.50 &~~ 0.70\\
 $|\downarrow\rangle$ &~~&  $0.01i$ &~~ --0.08 &~~ 0.07 &~~ --0.07$i$ &~~ 0.08$i$ \\\hline
\end{tabular}
\end{center}
\label{table2}
\end{table}

\section{Summary}
Materials such as KOsO$_4$ with an $e_g^1$ configuration are expected to have
a negligible orbital moment. Mixing of $t_{2g}$ character is required,
which is aided by
small crystal field splitting and structural symmetry lifting.
We have studied the interplay of strong correlation
effects and large spin-orbit coupling strength, and have found that an additional
characteristic is very important: the additional symmetry breaking of the
electronic state 
by spin-orbit
coupling itself. 
The spin-direction dependent orbital moment in this Os$^{7+}$ $e_g^1$ system
has been analyzed and understood. The occupied orbital without spin-orbit
coupling is $d_{z^2}$$|\uparrow\rangle$. For spin along the [001] axis, indeed
there is negligible mixing with
$m_l\neq$ 0 orbitals and the only change due to SOC is a few percent reduction in
the spin moment. 

For the spin along the in-plane [100] axis, however, SOC further breaks 
$x\leftrightarrow y$ symmetry, inducing 
a population imbalance in the $m_l=-2$ and $m_l=+2$ orbitals  
relative to the spin direction, which drives 
the unexpectedly large orbital moment $M_L$ = --0.2$\mu_B$.
This moment cancels half of the Os spin moment, and the accompanying
magnetocrystalline anisotropy
favors this [100] spin orientation. 

\section{Acknowledgments}
We acknowledge J. Yamaura for communications on resistivity measurement, 
and J. Kune\v{s} for useful discussions on the calculations of
Wannier functions including SOC.
This research was supported by National Research Foundation of Korea 
Grant No. NRF-2013R1A1A2A10008946 (K.W.L.)
and by U.S. Department of Energy Grant No. DE-FG02-04ER46111 (W.E.P.).

\end{document}